\documentclass[reprint,twocolumn,superscriptaddress,aps]{revtex4-1}
\usepackage{graphicx}
\usepackage{amsmath}
\usepackage{latexsym}
\usepackage{amsfonts}
\usepackage{amssymb}
\usepackage{array}
\usepackage{epsfig}
\usepackage{txfonts}
\usepackage[colorlinks=true,linkcolor=blue,urlcolor=blue,citecolor=blue]{hyperref}
\usepackage{color}
\usepackage{hyperref}
\usepackage{braket}

\setcitestyle{numbers,square,citesep={,\kern-2pt}}

\begin{document}

\setlength{\abovedisplayskip}{4pt}
\setlength{\belowdisplayskip}{4pt}

\frenchspacing
\spaceskip 2pt plus 0.5pt minus 0.5pt

\hyphenpenalty=10000
\exhyphenpenalty=5000

\title{Magnetic phases of spin-1 lattice gases with random interactions}

\author{Kenneth D. McAlpine}
\affiliation{Centre for Theoretical Atomic, Molecular, and Optical Physics, School of Mathematics and Physics, \\
Queen's University Belfast, Belfast BT7 1NN, United Kingdom}
\author{Simone Paganelli}
\affiliation{Dipartimento di Scienze Fisiche e Chimiche, Universit\`a dell'Aquila, via Vetoio, I-67010 Coppito-L'Aquila, Italy}
\author{Sergio Ciuchi}
\affiliation{Dipartimento di Scienze Fisiche e Chimiche, Universit\`a dell'Aquila, via Vetoio, I-67010 Coppito-L'Aquila, Italy}
\affiliation{Consiglio Nazionale delle Ricerche (CNR-ISC) Via dei Taurini, I-00185 Rome, Italy}
\author{Anna Sanpera}
\affiliation{ICREA, Pg.\,Llu\'is Companys 23, E-08010 Barcelona, Spain}
\affiliation{Departament de F\'{i}sica, Universitat Aut\`{o}noma de Barcelona, E-08193 Bellaterra, Spain}
\author{Gabriele De Chiara}
\affiliation{Centre for Theoretical Atomic, Molecular, and Optical Physics, School of Mathematics and Physics, \\
Queen's University Belfast, Belfast BT7 1NN, United Kingdom}

\begin{abstract}
A spin-1 atomic gas in an optical lattice,  in the unit-filling Mott Insulator (MI) phase and  in the presence of disordered spin-dependent interaction, is considered. In this regime, at zero temperature, the system is well described by a disordered rotationally invariant spin-1 bilinear-biquadratic model. We study, via the density matrix renormalization group algorithm,  a bounded disorder model such that the spin interactions can be locally either ferromagnetic or antiferromagnetic. Random interactions induce the appearance of a disordered ferromagnetic phase characterized by a nonvanishing value of the spin glass order parameter across the boundary between a ferromagnetic phase and a dimer phase exhibiting random singlet order. We also study the distribution of the block entanglement entropy in the different regions.
\end{abstract}
\date{\today}
\maketitle

\section{\hspace{-6pt}Introduction}
\vspace{-12pt}
The  interest for quantum magnetism has grown recently together with the possibility to simulate a huge variety of
spin models with ultracold atoms trapped in optical lattices \cite{Jaksch2005,bookLewenstein}.
The magnetic properties of the gas derive from their  low-energy hyperfine level structure which provides an extra spin degree of freedom ${F=I+J}$, where
$I$ is the nuclear spin and $J$ is the total electronic angular momentum.
Several experiments have been realized with spinor atomic gases \cite{KawaguchiUeda,StamperKurnRMP},  especially with alkali atoms, having  electronic angular momentum  $1/2$ and two hyperfine levels with ${F=\left| I\pm1/2\right|}$.  Here, we are interested in bosonic particles with lower energy hyperfine manifold corresponding to $F=1$; this is the case, for example, of $^{23}\textrm{Na}$,  $^{87}\textrm{Rb}$, $^{41}\textrm{K}$, and $^{7}\textrm{Li}$ \cite{StamperKurnRMP}.

A system of bosons trapped in a deep optical lattice potential is well described by a Bose-Hubbard model, consisting of two competing terms: the hopping
between lattice sites and the repulsive interaction produced by local  scattering.
If the spin orientation is fixed by an external magnetic field, as happens when the gas is confined in a magnetic trap, a scalar model is sufficient to describe the system \cite{Fisher89,Freericks1994}. Conversely, if the spin orientation is not externally constrained, as in the case of optical trapping, the spinor character of the gas has to be taken into account.

Bosonic interactions, due to two-body $s$-wave collisions, are sensitive to the spin degree of freedom and contribute to the ordering properties at zero temperature.
For spin-1 particles, two possible scattering channels open, one with total spin ${s=0}$ and one with total spin ${s=2}$, corresponding respectively  to a scattering length $a_0$ and $a_2$.  This leads to a spin-dependent interaction \cite{Imambekov03,Tsuchiya04,Rizzi05} which, depending on the scattering lengths $a_0$ and $a_2$, can be ferromagnetic or antiferromagnetic.
Spin-1 condensates have been observed in Refs. \cite{Stamper-Kurn1998,Stenger1998} and the two-body interaction has been studied in Refs. \cite{Ohmi98,Ho98,Law98}.

At zero temperature, the phase diagram of interacting spin-1 bosons on a lattice consists of the Mott insulator (MI) lobes as in the scalar case while for smaller interactions the system is superfluid (SF) \cite{Rizzi05}.
In the MI phase, density fluctuations are suppressed while the spin degrees of freedom give rise to a variety of possible magnetic phases.
In order to explore the magnetic  behavior  inside the MI phase, it is possible to map the problem into a spin Hamiltonian by a perturbative expansion in the hopping parameter.  For unit filling, the MI phase has one boson per site, so the Bose-Hubbard spin-1 chain is mapped onto the bilinear-biquadratic spin-1  Hamiltonian \cite{Fath91,Fath95,Schollwock96,Buchta05,DeChiaraAug11,Lauchli2006,Moreno-Cardoner2014a}, preserving the SO(3)  symmetry of the hyperfine atomic splitting.

Among all the opportunities ultracold atoms can offer, there is the possibility to produce a disordered potential in a controlled way by extra speckle potentials \cite{Horak98,Boiron99}, optical superlattices  of incommensurate frequencies \cite{Dinier01,Roth03,Damski03,Fallani2007,Deissler2010,Deissler2011a,Habibian2013,Habibian2013a}, holographic masks \cite{Choi1547, Zupancic}, or adding different atomic species randomly trapped in sites distributed across the sample and acting as impurities \cite{Gavish05,Massignan06,stasinska2012}.
In the so-called dirty bosons problems, the competition between the local disorder and the interactions gives rise to a Bose glass (BG) between the MI and SF  which exhibits localization but, unlike the MI, is compressible and gapless. The phase diagram in the presence of this type of disorder has been studied both for spinless \cite{Fisher89,Gurarie09,Bissbort10,Damski03,Lewenstein07,Lugan2008} and for spinful bosons \cite{acki2011,Paganelli2011,Warnes2012,Wagner2012,Kimura2013,StamperKurnRMP,Nabi2016}.

On the other hand, disorder can  also be present in the interaction terms of the Hamiltonian. Such a possibility is more difficult to control and reproduce experimentally since it requires, for example, varying the scattering length of the $s$-wave collision at random. Recently, the Bose-Hubbard model in the presence of random interactions has been analyzed numerically in the context of many-body localization \cite{sierant2016many}.
 Magnetic Feshbach resonances can be exploited for this purpose \cite{Chin10,Gimperlein05}; nevertheless, for spinor bosons, an external  magnetic field would freeze the spin degrees of freedom and it would map the system into a spinless one.
However, from a theoretical point of view, disorder in the interaction is especially interesting for spin-1 bosons, where different spin phases emerge inside the MI phase, contrary to the spinless case, where the BG separates MI to SF.
 Therefore, in order to implement random local interactions  without freezing the spin degrees of freedom, different techniques are in order.  Adopting optical Feshbach resonances could be a possible route.
This technique  employs a laser tuned near  a photoassociation transition with an excited molecular level. The method was developed theoretically \cite{Fedichev96,Bohn97,Papoular10,Nicholson2015}   and has been proposed to quantum simulate frustrated magnetism with spinor Bose gases in 2D lattices\cite{Debelhoir16}. Experimental observations of optical Feshbach resonances have recently been carried out with alkali-metal atoms \cite{Fatemi2000,Theis2004} and  alkali-earth-metal-like  atoms \cite{Enomoto2008,Yamazaki2010,Blatt2011, Yan2013}, where atom losses have been shown to be reduced  as suggested in Ref. \cite{Ciurylo2005}. Nevertheless, atomic losses are still limiting the lifetime of the gas below the millisecond regime and novel ways to address Feshbach resonances for spinor condensates are highly desirable.

In this paper, we study the case of a spin-1 system in the MI phase with one boson per site, when disorder mixes locally  ferromagnetic and antiferromagnetic interactions.  By mapping the spin-1 Bose-Hubbard system into the corresponding spin model (bilinear-biquadratic), we find numerically the ground state of the system for each disorder realization
 using the density matrix renormalization group (DMRG) \cite{White92, DeChiara08}.
 To identify the possible phases of the model, we analyze the ferromagnetic and the dimer order parameters, together with a normalized Edwards-Anderson (EA) order parameter \cite{Albino00}.
Our results indicate that the phase diagram consists of three phases: a ferromagnetic phase, a random singlet phase with a nonzero dimer order parameter, and an intermediate phase between the first two phases.
 This intermediate phase exhibits nonvanishing EA parameter and we argue that it can be identified as a disordered ``large-spin'' regime \cite{Quito15}.
By studying the entanglement entropy  between two half chains, we exclude the existence of random singlet ordering in the large-spin phase.
A linear scaling with the system size of the ferromagnetic domain walls suggests that the intermediate phase could be a locally disordered ferromagnet containing microscopic magnetized droplets.

The paper is organized as follows: In Sec. \ref{sec:model}, we describe the spin-1 Bose-Hubbard model and the corresponding bilinear-biquadratic spin-1 model for the MI phase and unit density. Starting from disordered two-body atomic interactions, we derive the corresponding random coefficients of the spin model. In Sec. \ref{sec:numerical}, we present numerical results for local order parameters, showing the existence, together with the ferromagnetic and the dimer phases, of a large-spin phase. In Sec. \ref{sec:entropy}, we analyze the entanglement properties of the system by studying the block entanglement entropy.  In Sec. \ref{sec:coex},  we study the scaling of the domain walls with the system size, and finally in Sec. \ref{sec:conclusions}, we summarize and present open questions.

\vspace{-12pt}
\section{\hspace{-6pt}Model}\label{sec:model}
\vspace{-12pt}
We consider a one-dimensional chain of spin-1 bosons in an optical lattice. The effective Bose-Hubbard Hamiltonian for spin-1 bosons is \cite{Imambekov03,Rizzi05}
\begin{equation}\label{Bose}
\begin{aligned}
{H}&=\frac{U_0}{2}\sum_i{n}_i\left({n}_i-1\right)+\frac{1}{2}\sum_i U_{2i}\left({\mathbf{S}}_i^2-2{n}_i\right)-\mu\sum_i{n}_i\\
&-t\sum_{i,\sigma}\left({a}_{i,\sigma}^\dagger{a}_{i+1,\sigma}+{a}_{i+1,\sigma}^\dagger{a}_{i,\sigma}\right)
\end{aligned}
\end{equation}
with ${a}_{i,\sigma}^\dagger$ and ${a}_{i,\sigma}$ being the creation and annihilation operators for site $i$, spin component $\sigma=0,\pm 1$, respectively,
\begin{equation}
{n}_i=\sum_\sigma{a}_{i,\sigma}^\dagger{a}_{i,\sigma}
\end{equation}
is the total numbers of particles on site $i$, and
\begin{equation}
{\mathbf{S}}_i=\sum_{\sigma,\sigma'}{a}_{i,\sigma}^\dagger{\mathbf{T}}_{\sigma,\sigma'}{a}_{i,\sigma'}
\end{equation}
the total spin on site $i$ with ${\mathbf{T}}_{\sigma,\sigma'}$ the spin-1 angular momentum matrix elements.
The on-site interactions are described by the first two terms of Eq. \eqref{Bose}. The local density coupling is assumed to be  site independent  with coupling constant $U_0$. The second interaction term takes into account the local spin and we assume its coupling constant $U_{2i}$ to be disordered and site dependent.
The third term is an on-site energy, with $\mu$ being the chemical potential, and finally
the fourth term of \eqref{Bose} is the tunneling term describing hopping
between nearest-neighbor sites with tunneling amplitude $t$.
Our focus is on the strong interaction regime ${(t\ll U_0)}$; further, we set $U_0$ as the energy scale unit ${(U_0=1)}$.
Our analysis is restricted to the first Mott lobe with unit filling factor that extends, as it is easy to see in the atomic limit, in the interval ${U_2 \in(-1;0.5)}$ \cite{acki2011,Paganelli2011}.

By considering a spin-1 chain of $L$ sites with open boundary conditions, deep in the Mott insulator, with strong interactions and unit filling per site, ${{n}_i=1}$, the Hamiltonian of Eq. \eqref{Bose} can be mapped into a spin-1 bilinear-biquadratic Hamiltonian:
\begin{equation}\label{Hamiltonian}
H=\sum_{i=1}^{L-1}H_i=\sum_{i=1}^{L-1}[J_{1i}\left(\mathbf{S}_i\cdot\mathbf{S}_{i+1}\right)+J_{2i}\left(\mathbf{S}_i\cdot\mathbf{S}_{i+1}\right)^2],
\end{equation}
where ${\mathbf{S}_i=(S_{xi},S_{yi},S_{zi})}$ are the $i$th site angular momentum operators and ${J_{1i}=f\left(U_0,U_{2i},U_{2(i+1)}\right)}$ and ${J_{2i}=g\left(U_0,U_{2i},U_{2(i+1)}\right)}$ [see Appendix \ref{sec:derivation} for the derivation of this mapping] are the linear and quadratic exchange coefficients between spin sites $i$ and $i+1$.

For fixed $U_0$, distinct values of $U_{2i}$ and $U_{2(i+1)}$ in the region of interest correspond to a unique value of the ratio $J_{2i}$ to $J_{1i}$, which is conveniently parametrized by the angle $\theta_i$ defined as
\begin{equation}\label{U2Theta}
\theta_i=\arctan\left(\frac{J_{2i}}{J_{1i}}\right)-\pi.
\end{equation}

\vspace{-12pt}
\subsection{\hspace{-6pt}Homogeneous coupling constants}
\vspace{-12pt}
In the case of homogeneous exchange constants ${(U_{2i}=U_2, \,\forall \,i)}$, the Hamiltonian of Eq. \eqref{Hamiltonian} becomes the general bilinear-biquadratic Hamiltonian \cite{Fath91,Fath95,Schollwock96,Buchta05,Lauchli2006,DeChiaraAug11}:
\begin{equation}\label{eqn:bil-biq}
H_{BB}=J\sum_{i=1}^{L-1}[\cos\theta\left(\mathbf{S}_i\cdot\mathbf{S}_{i+1}\right)+\sin\theta\left(\mathbf{S}_i\cdot\mathbf{S}_{i+1}\right)^2],
\end{equation}
where $J_1$ and $J_2$ have been replaced using Eq. \eqref{U2Theta} and ${J=\sqrt{J_1^2+J_2^2}}$.

The ground-state phase diagram for the model with $H_{BB}$ is as follows: For ${\pi/2<\theta\le\pi}$ and ${-\pi\le\theta<-3\pi/4}$, the system is in the ferromagnetic phase characterized by a net spontaneous magnetization along a direction which breaks spontaneously space isotropy. For ${-3\pi/4<\theta < -\pi/4}$, the system is in the dimerized phase, characterized by a nonvanishing dimer order parameter. In this phase, translational invariance is broken and imperfect singlets of neighboring spins appear. For ${-\pi/4<\theta<\pi/4}$, the system is in the gapped Haldane phase exhibiting absence of any local order, a nonzero string order parameter and entanglement spectrum with even number of degeneracies. Finally, for ${\pi/4<\theta<\pi/2}$, the system is in the so-called critical phase characterized by quasi-long-range quadrupolar order.

The region of interest for ultracold atoms, ${U_2\in(-1;0.5)}$, corresponds thus to the region ${\theta\in(-\pi+\arctan(1/3);-\pi/2)}$ which encompasses part of the ferromagnetic phase, part of the dimer phase and the transition between them, occurring at $U_2=0$. The full phase diagram could be realized by preparing the highest energy state of 1D optical lattices or by using spin ladders \cite{Garcia04}.

The phase transition that occurs at ${U_2 = 0}$, which corresponds to ${\theta=-3\pi/4}$, is the focus of this paper. Chubukov conjectured \cite{Chubukov91} that the ferromagnetic and dimerized phases were separated by a possible spin nematic phase, but due to an abnormal divergence in correlation length this transition was extremely hard to categorize. The existence of the nematic phase was widely studied numerically \cite{Fath95,Buchta05,Lauchli2006,Rizzi05,Oriol} and it is now accepted that the extension of the nematic phase is extremely small, suggesting a direct first-order transition between the ferromagnetic and dimer phase.

\vspace{-12pt}
\subsection{\hspace{-6pt}Disordered coupling constants}
\vspace{-12pt}
We implement disorder in the spin system by considering random exchange coefficients $J_{1i}$ and $J_{2i}$ obtained by random interaction constants $U_{2i}$.
The resulting Hamiltonian has the form
\begin{equation}
H_{BB}=\sum_{i=1}^{L-1}J_{i}(\cos\theta_{i}(\mathbf{S}_{i}\cdot\mathbf{S}_{i+1})+\sin\theta_{i}(\mathbf{S}_{i}\cdot\mathbf{S}_{i+1})^{2}),
\end{equation}
with ${J_{i}=\sqrt{J_{1i}^{2}+J_{2i}^{2}}}$. In this work, disorder is confined to a range around a central point, $U_{2C}$, of width $\eta$.  Thus $U_{2i}$ is generated for $i=1, 2, \dots, L$ via
\begin{equation}\label{U2Gen}
U_{2i}=U_{2C}+\eta(2\zeta_i-1),
\end{equation}
where $\zeta_i$ is a uniformly distributed random variable between $0$ and $1$, leading to $U_{2i}$ being uniformly distributed between ${[U_{2C}-\eta,U_{2C}+\eta]}$.
To enforce locally the constraint  ${U_2\in(-1;0.5)}$, we choose $U_{2C}\in(-1+\eta;0.5-\eta)$.
Throughout this paper, the value of $\eta$ is fixed at $\eta=0.1$.
A situation where disorder appears mostly on $U_2$ can be realized by considering
anticorrelated fluctuations of the scattering lengths.
Since the coupling constants depend on the $s$-wave scattering lengths $a_s$ $(s=0,2)$ as $U_0\propto a_0 + 2 a_2$  and $U_2\propto a_2- a_0$ (details are given in Refs. \cite{Ohmi98,Ho98,Law98}), it is sufficient to consider a fluctuation $\delta_i$ in the scattering lengths such that  ${a_2^{(i)}=a_2+\delta_i}$ and ${a_0^{(i)}=a_0-2\delta_i}$.

\vspace{-12pt}
\section{\hspace{-6pt}Numerical Results}\label{sec:numerical}
\vspace{-12pt}
Results were obtained using the finite-size DMRG with open boundary conditions; 500 random configurations of exchange coefficients were used. In order to reduce stability issues related to the degeneracy of the ground state, a weak magnetic field $(-10^{-5}S_{z1})$ was added to the first spin in the chain in order to promote one of the almost degenerate ground states. It is important to stress that even if the total angular momentum commutes with the random bilinear-biquadratic Hamiltonian Eq. \eqref{Hamiltonian}, the minimum energy state can assume values, between $-L$ and $L$ of the projection of the total angular momentum ${M_z =\sum_i S_{zi}}$, depending on the disorder configuration. In the DMRG simulations we used five complete sweeps and kept 80 states during the renormalization procedure corresponding to a maximum discarded weight of $10^{-5}$.
For all figures presented, the results obtained from a Hamiltonian with homogeneous exchange coefficients are labeled H and those with disordered exchange coefficients are labeled D.

\vspace{-12pt}
\subsection{\hspace{-6pt}Local magnetization and nearest-neighbor correlations}
\label{sec:local}
\vspace{-12pt}
\begin{figure}[t]
\centering
\includegraphics[trim=8.5cm 2.5cm 10cm 1.5cm, clip=true, width=0.95\columnwidth]{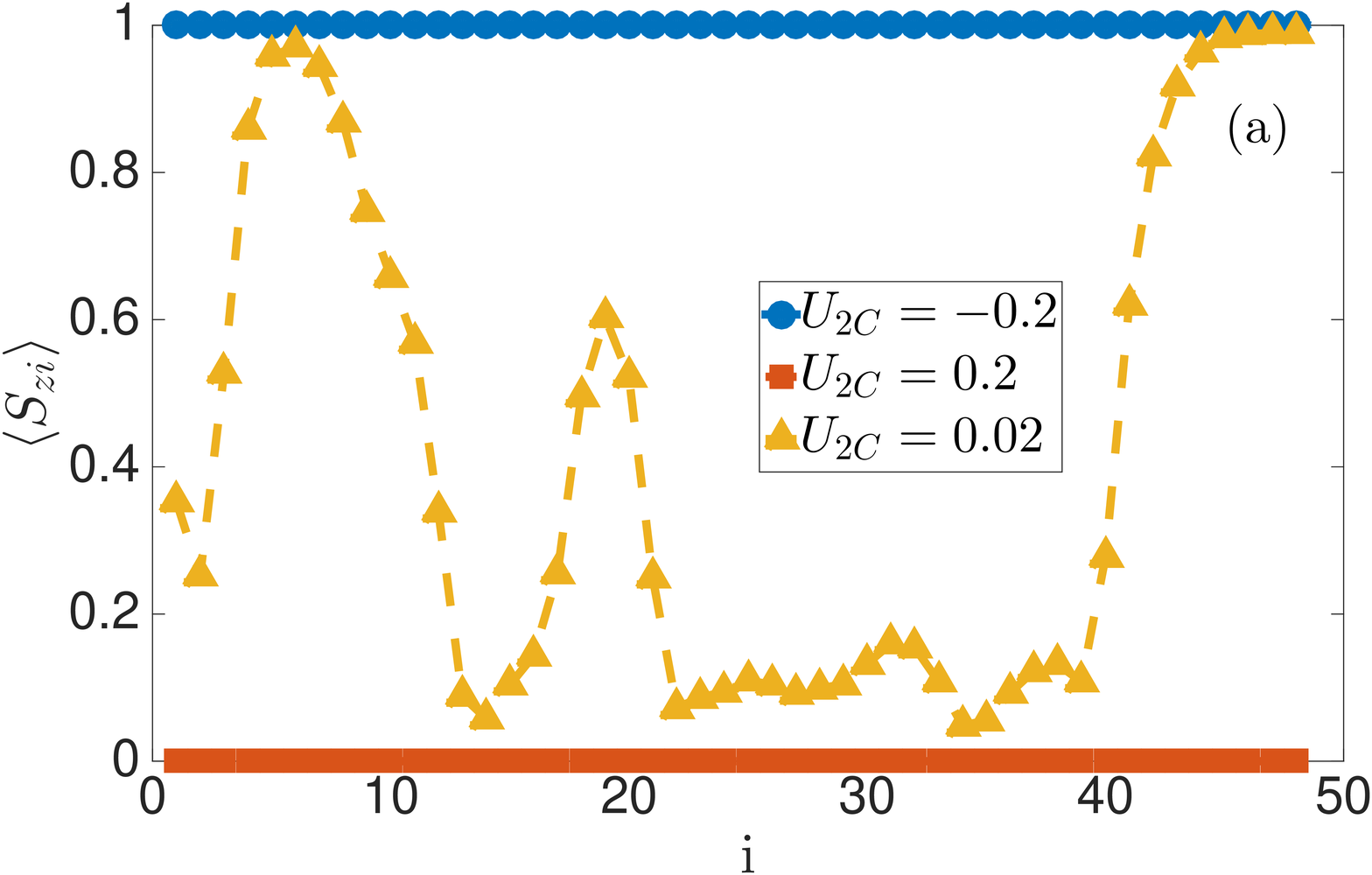}\\
\includegraphics[trim=8.5cm 0.45cm 10cm 1.5cm, clip=true, width=0.95\linewidth]{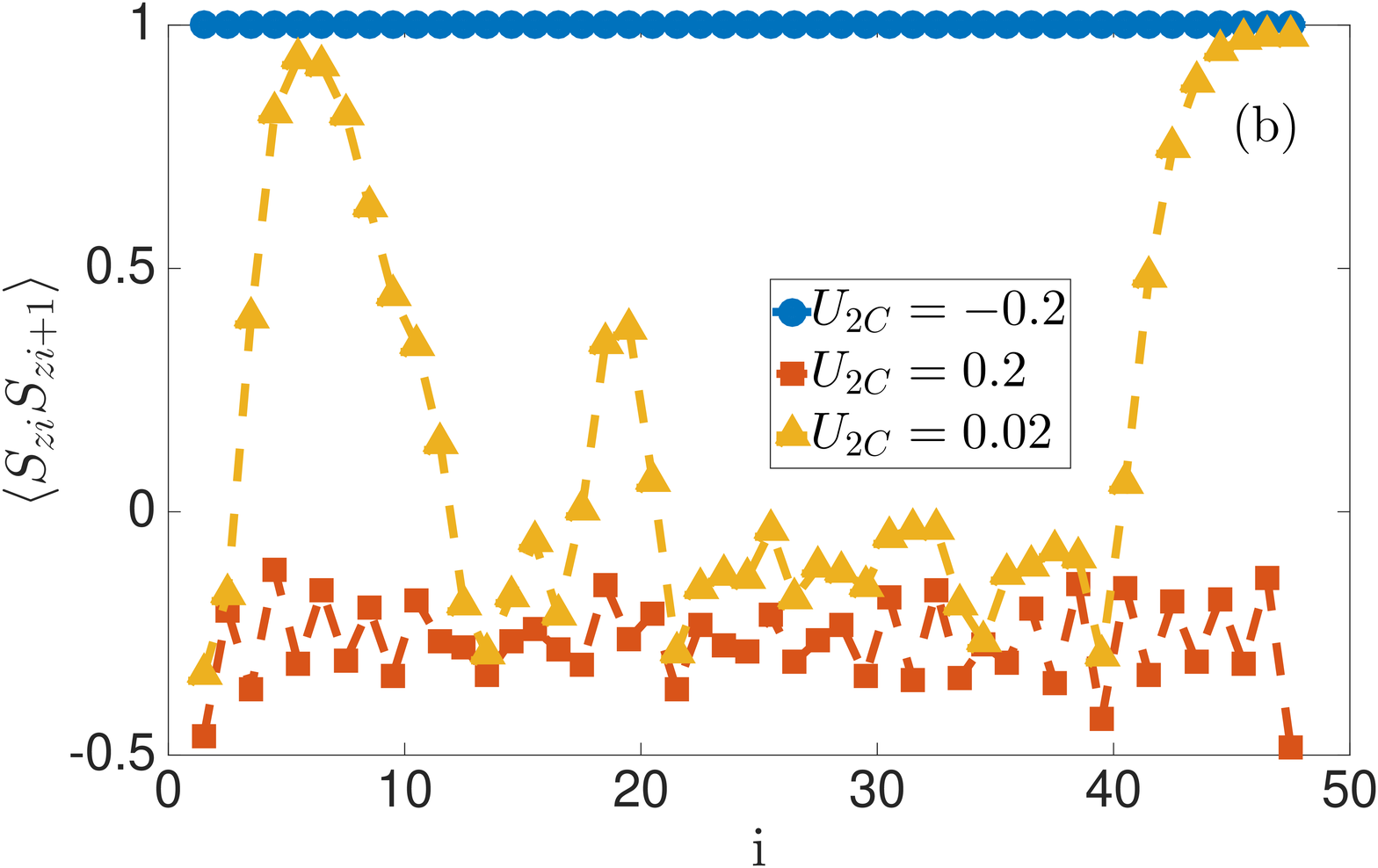}
\caption{Correlations for disordered spin-1 chains of length ${L=48}$, fully in the ferromagnetic regime (circles), fully in the dimerized regime (squares), and close to the transition point (triangles). (a) Local correlation $\langle{S}_{zi}\rangle$. (b) Nearest-neighbor correlation $\langle{S}_{zi}{S}_{z{i+1}}\rangle$. Diagrams plotted against chain site $i$.}
\label{fig:Examples}
\end{figure}

We start our analysis by computing the local magnetization $\langle S_{zi} \rangle$ and the nearest-neighbor spin-spin correlations $\langle S_{zi} S_{zi+1}\rangle$, shown in Fig. \ref{fig:Examples} for a single configuration of $\{\zeta_i\}$. For ${U_{2C}=-0.2}$, the system is deep in the ferromagnetic phase and the local magnetization is everywhere equal to $+1$ (thanks to the negative-symmetry-breaking field mentioned above), while the  correlations are exactly $1$, showing perfect spin alignment. For ${U_{2C}=0.2}$, the system has zero magnetization and an overall negative, thus antiferromagnetic, nearest-neighbor correlations. The plot also shows an even-odd effect that can be explained, as we show in the next section, assuming that the ground state is dimerized. Finally for ${U_{2C}=0.02}$ there is an intermediate behavior: The system is not fully magnetized but exhibits islands of nonzero magnetization with large nearest-neighbor correlations  surrounded by regions of very small magnetization with overall negative nearest-neighbor correlations. This is our first evidence of a disordered  intermediate region between the dimer and ferromagnetic phases.

\vspace{-12pt}
\subsection{\hspace{-6pt}Dimerization versus ferromagnetism}\label{DimerSection}
\vspace{-12pt}
To characterize the dimer phase, a common choice is the local dimer order parameter \cite{Fath95,Buchta05}:
\begin{equation}
D=[\langle H_i  - H_{i+1}\rangle ]_D,
\end{equation}
where $\langle\,\rangle$ is the quantum ground state average and $[\,]_D$ is the disorder average.
Because of the disordered nature of the chain we consider, we prefer to use a similar dimer order parameter $D_\epsilon$ defined in Ref. \cite{DeChiaraFeb11}
\begin{equation}\label{DimerLauchliDeChiara}
D_\epsilon=-\frac{1}{L}\sum_{mn}\sin\left[\frac{\pi}{2}(m+n)\right][G_z(m,n)]_D,
\end{equation}
where $G_z(m,n)$ is the two-point correlation function:
\begin{equation}
G_z(m,n)\equiv\langle{S}_{z_m}{S}_{z_n}\rangle - \langle{S}_{z_m}\rangle\langle{S}_{z_n}\rangle.
\end{equation}
While for the homogeneous model $D$ and $D_\epsilon$ have been shown to be equivalent \cite{DeChiaraFeb11}, for the model with disorder, $D_\epsilon$ converges faster thanks to a self-averaging effect due to the summation on all sites. Moreover, $D_\epsilon$ can be observed experimentally with a quantum nondemolition measurement based on the Faraday effect \cite{DeChiaraFeb11}.

\begin{figure}[t] 
\centering
\includegraphics[trim=7cm 0.45cm 13.5cm 0.5cm, clip=true, width=0.475\columnwidth]{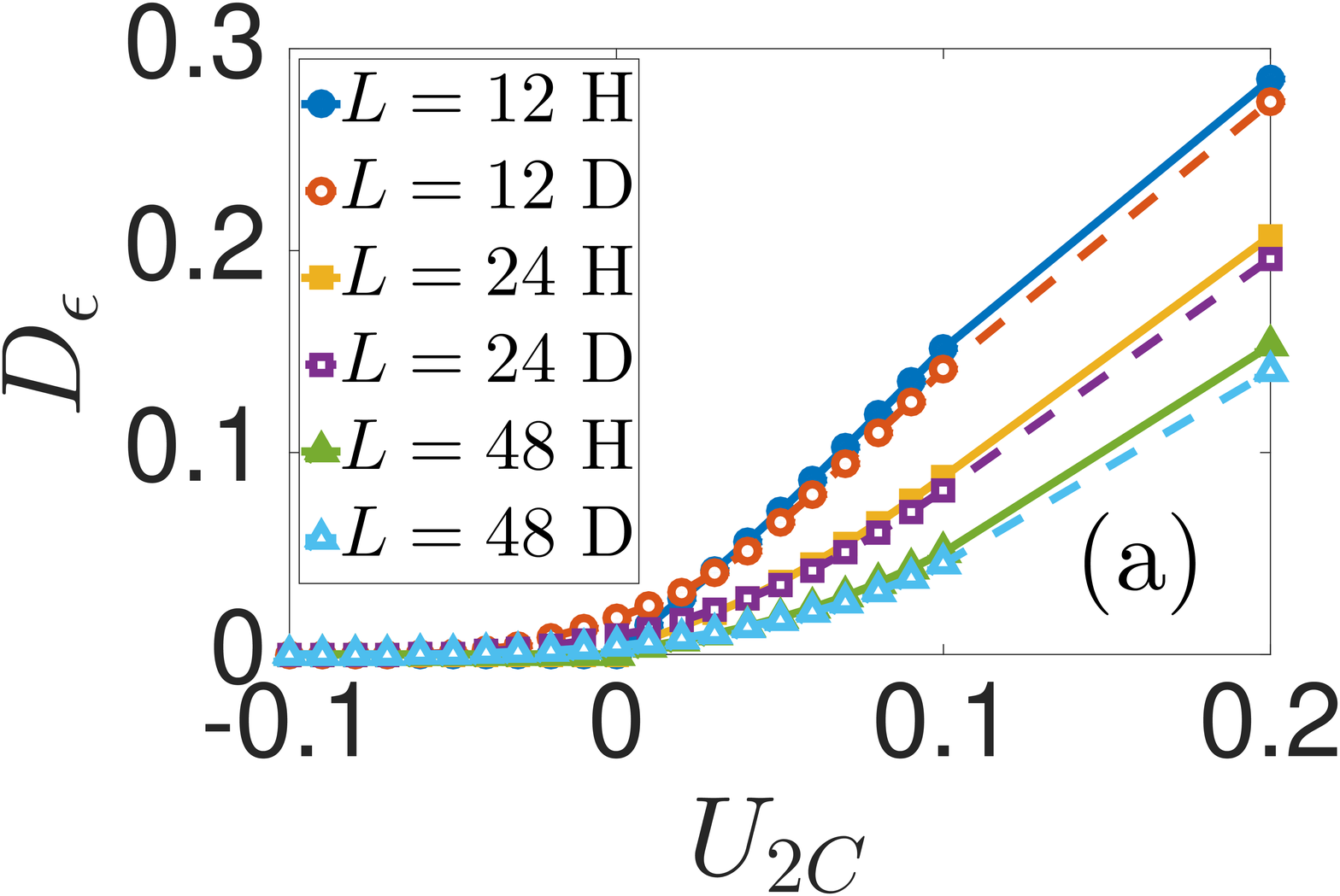}
\includegraphics[trim=7.cm 0.45cm 13.5cm 0.5cm, clip=true, width=0.475\columnwidth]{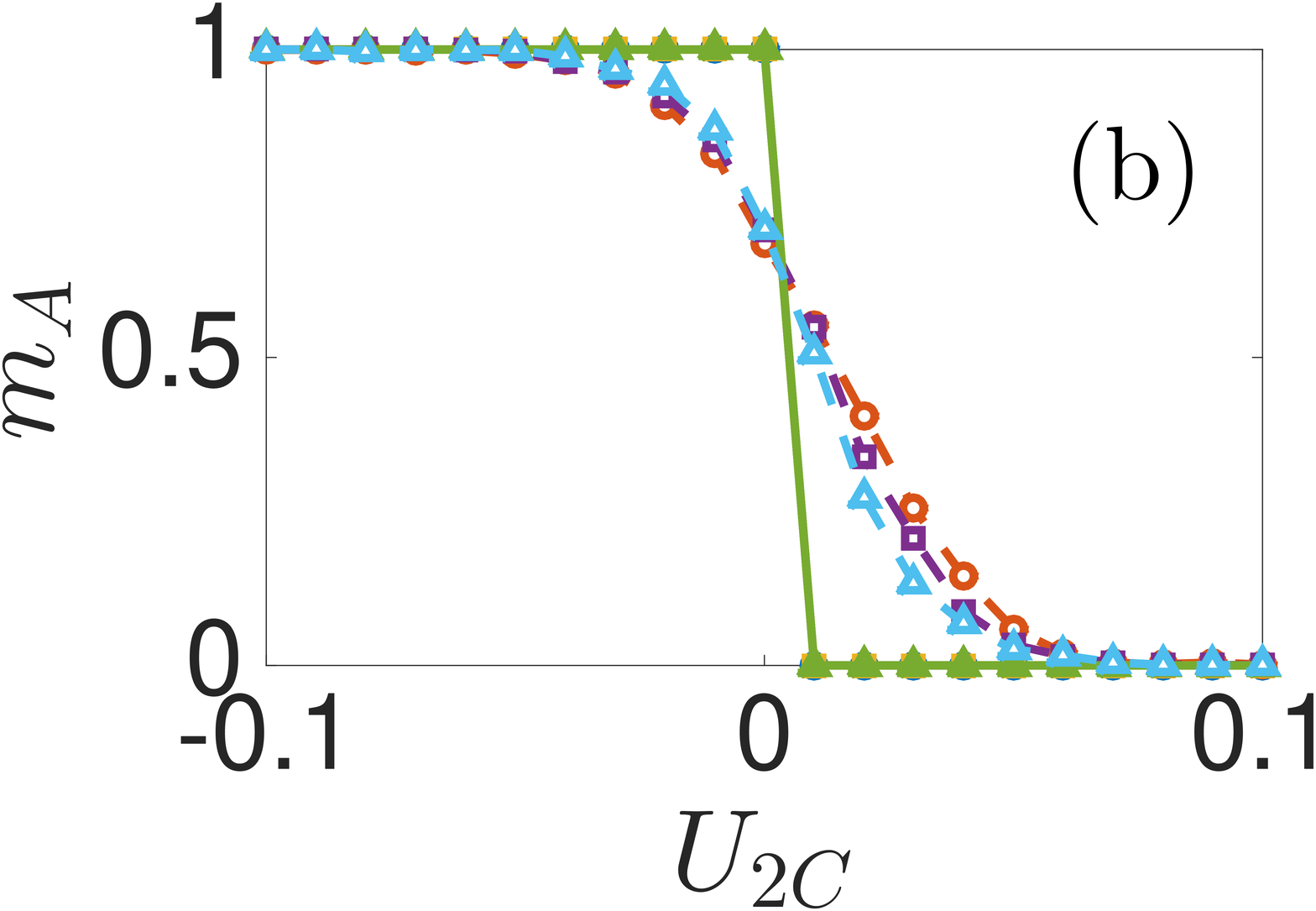}
\includegraphics[trim=10.5cm 0.3cm 7cm 0.15cm, clip=true, width=0.95\columnwidth]{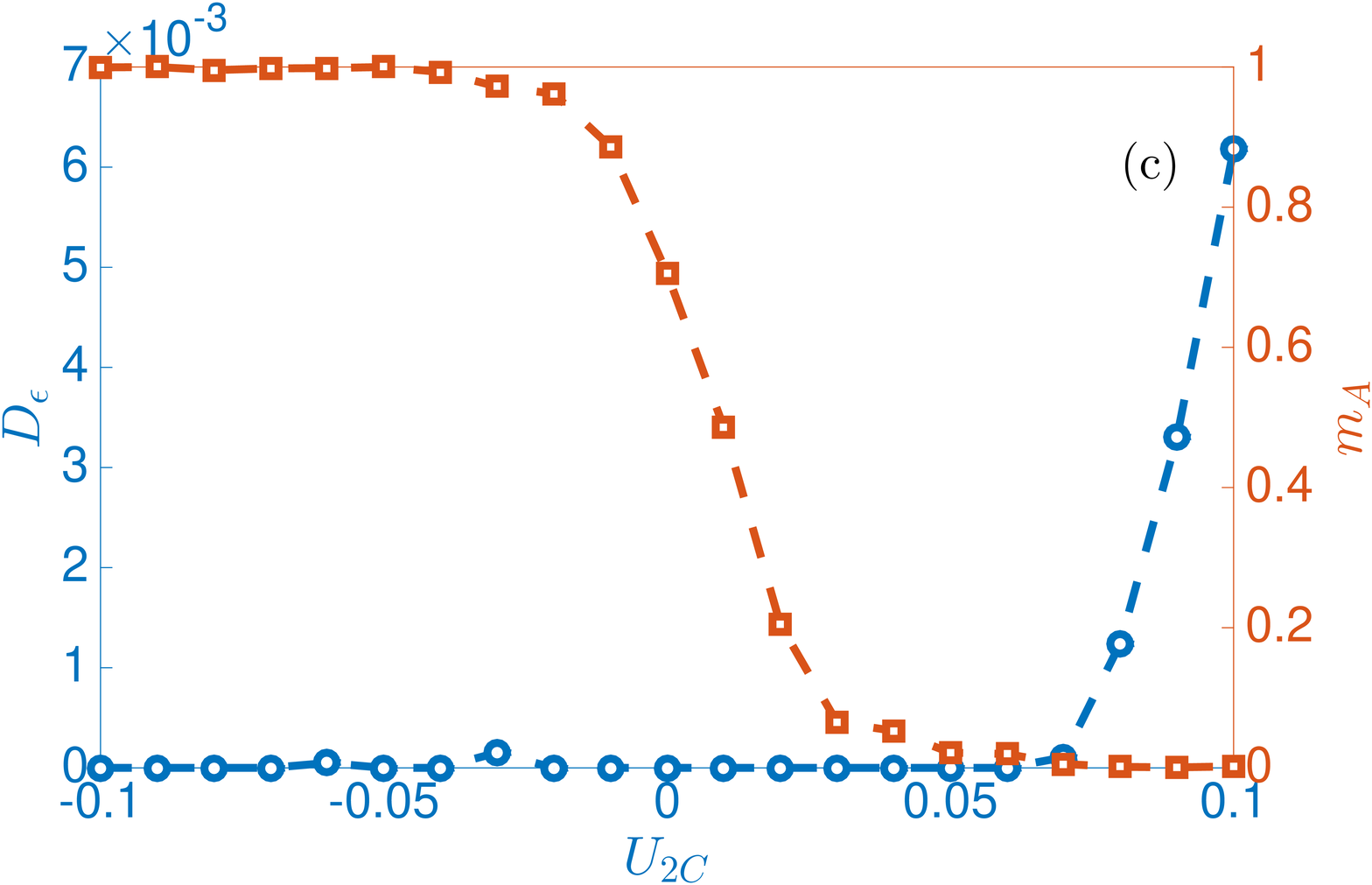}
\caption{(a) Dimer order parameter $D_\epsilon$ against $U_{2C}$ for differing lengths $L=12,24,48$. The label $H$ is for the homogeneous system (filled symbols) and $D$ is for the disordered case (empty symbols). The lines (solid for homogeneous and dashed for the disordered model) connect the points and are only a guide to the eyes. (b) Average magnetization $m_A$ along the $z$-axis; color codes are the same as in panel (a). (c) Dimer order parameter and magnetization of the disordered systems extrapolated to infinite length.
}
\label{fig:Dimer}
\end{figure}

As is customary, the average magnetization along the $z$ axis is defined as
\begin{equation}
\label{Magnetization}
m_A=\frac{1}{L}\sum_i[\langle S_{zi}\rangle]_D.
\end{equation}

Figure \ref{fig:Dimer}(a) shows the average dimer order parameter and Fig. \ref{fig:Dimer}(b) shows the average magnetization of the spin-1 chain throughout the region of interest for both the homogeneous and disordered cases.
For short lengths, we observe nonzero dimerization below the first-order transition point for the homogeneous model $(U_{2C}=0)$. Disorder slightly suppresses the dimer order parameter, which has a fairly large finite-size effect. On the other hand, magnetization does not show strong finite-size effects, but it is considerably affected by the presence of disorder which rounds the sudden jump observed in the homogeneous case. In Fig. \ref{fig:Dimer}(c), we show the results for the average dimer order parameter and magnetization after a finite-size extrapolation to infinite lengths (see details in Appendix \ref{SupFigs}). In contrast to the homogeneous case, the results clearly show that the magnetization is continuous at $U_{2C}=0$.

\begin{figure}[b] 
\centering
\includegraphics[trim=8.5cm 0.45cm 10cm 1.5cm, clip=true, width=0.95\columnwidth]{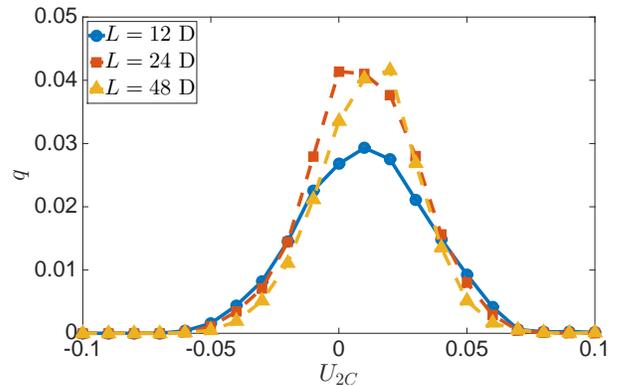}
\caption{Edwards-Anderson order parameter $q$ against $U_{2C}$ for the disordered  system for $L=12,24,48$. Lines are only a guide to the eye.}
\label{fig:Qprime}
\end{figure}
\vspace{-12pt}
\subsection{\hspace{-6pt}Spin glass order parameter in the intermediate phase}
\vspace{-12pt}
To ascertain the nature of the intermediate phase, we compute the Edwards-Anderson order parameter, defined as \cite{EdwardsAnderson}
\begin{equation}
      \label{eq:q}
      q=\frac{1}{L}\sum_i[\langle {S}_{zi}\rangle^2]_D-[\langle {S}_{zi}\rangle]^2_D.
\end{equation}
The term subtracted in Eq. \eqref{eq:q} takes into account the possibility of finite magnetization in the disordered magnetic phase.
It is clear that this parameter is zero in the homogeneous case and takes into account correlations of the local magnetization with the disorder variables.
This parameter is shown in Fig. \ref{fig:Qprime} for different chain lengths. The spin glass order parameter is not affected by finite-size effects and its behavior shows that the maximum of correlations between local magnetization and disorder occurs for $U_{2C}>0$. From the Edwards-Anderson order, parameter we infer a range for the intermediate phase of $-0.051\lesssim U_{2C}\lesssim0.073$.

The question thus arises as to whether the intermediate phase that spans from the vanishing of the ferromagnetic phase to the onset of the dimer phase is a new phase that can be characterized, for example, by random singlet order or by disorder induced magnetic domains.

In a previous work, Quito {\it et al.} \cite{Quito15}, using the strong-disorder renormalization group (SDRG) \cite{Igloi05}, found a large-spin phase separating the ferromagnetic phase and the random singlet phase.
The term {\it large-spin} is in reference to the renormalization procedure to find the ground state of the spin chain.

Here we summarize the renormalization group procedure of Quito {\it et al.} \cite{Quito15}. The SDRG uses a decimation procedure similar to the procedure originally pioneered by Dasgupta and Ma \cite{Dasgupta80,Fisher1995}. Each decimation replaces the spin pair corresponding to the largest energy gap between the ground state and the first excited state of the local Hamiltonian, $H_i$, with a new effective spin. For a spin-1 chain, the Hamiltonian $H_i$ corresponds to a spin pair located at sites $i$ and $i+1$. If their coupling is ferromagnetic, i.e., $U_{2i},\,U_{2(i+1)} \in (-1;0)$, the pair of spins is replaced by a new effective spin of size spin 2 and new effective couplings are produced between the new spin and the spins on sites $i-1$ and $i+2$. For $U_{2i},\,U_{2(i+1)} \in (0;0.5)$, corresponding in the homogeneous model to the dimer phase, the two spins form a singlet. In the decimation procedure, a spin pair of this type is removed and a new effective coupling between the spins on sites $i-1$ and $i+2$ is found. Repeating this decimation procedure will eventually reduce a chain in which all $U_{2i}$ lie in the range $(-1;0)$ to a single effective spin of size $L$, whereas a chain in which all $U_{2i}$ lie in the range $(0;0.5)$ will reduce to an effective spin of size $0$. This large-spin phase describes a region in which a chain may have a final effective spin which lies somewhere between $0$ and $L$, indicating that it is neither fully ferromagnetic nor fully dimerized.

Our DMRG calculations also show that in the intermediate phase different instances of disorder exhibit a wide distribution of total magnetization, precisely as predicted in Ref. \cite{Quito15}.
This indicates that the intermediate phase could coincide with the large-spin phase found in Ref. \cite{Quito15}. It should be noted, however, that Quito {\it et al.} \cite{Quito15} modeled their disorder in such a way that the random coefficients $J_{1i}$ and $J_{2i}$ were independent of each other, while in this work $J_{1i}$ and $J_{i2}$ are both dependent on $U_{2i}$ and $U_{2(i+1)}$.
\vspace{-12pt}
\subsection{\hspace{-6pt}Entanglement entropy}\label{sec:entropy}
\vspace{-12pt}
Further to the three order parameters discussed previously, we consider the block entanglement entropy. The advantage of this bipartite entanglement measure is that as well as being discontinuous at a first-order transition \cite{Orus11,Malvezzi16} its spatial distribution in a disordered model can give insight into the bonds existing between blocks of spins. In the so-called random singlet phase, found for instance in the random coupling $XX$ model and in the random transverse-field Ising model for spin-1/2 \cite{Laflorencie2005,Binosi07}, the probability distribution of the entanglement entropy is characterized by pronounced peaks at integer values, indicating the occurrence of maximally entangled states of pairs of spins belonging each to one side of the partition. Very recently the scaling of the entanglement entropy distribution away from integer points has been studied in Refs. \cite{devakul2016probability}.
Finding the probability distribution of the entanglement entropy is particularly interesting for our problem because we can confirm the existence or absence of a random singlet phase in this model.
\begin{figure}[t] 
\centering
\includegraphics[trim=9cm 0.45cm 10cm 1.5cm, clip=true, width=0.95\columnwidth]{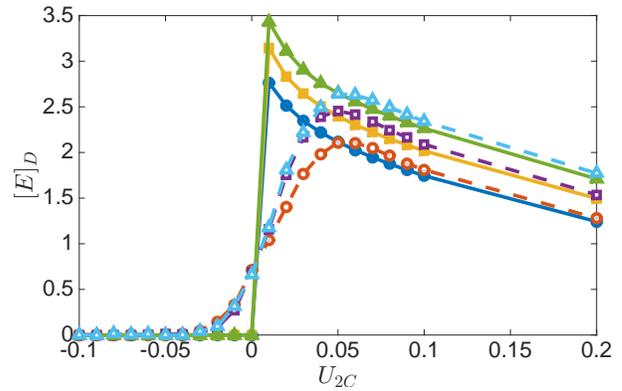}
\caption{Average von Neumann entropy $[E]_D$ against $U_{2C}$ for variable chain lengths with the same color-coding as in Fig. \ref{fig:Dimer}. The lines (solid for homogeneous and dashed for the disordered model) connect the points and are only a guide to the eyes.}
\label{fig:MVN}
\end{figure}
The block entanglement entropy measures the entanglement between two blocks of lengths $\ell$ and $L-\ell$ and can be calculated as the von Neumann entropy of the reduced density matrix of one of the two blocks \cite{Vidal2003}:
\begin{equation}
\label{eq:EE}
E=-\mathrm{Tr}\rho_\ell\log_2\rho_\ell
\end{equation}
with
\begin{equation}
\rho_\ell=\mathrm{Tr}_{L-\ell}|\psi_G\rangle\langle\psi_G|,
\end{equation}
where $|\psi_G\rangle$ is the ground state of Hamiltonian \eqref{Hamiltonian} and we set $\ell=L/2$.

In Fig. \ref{fig:MVN} we show the results for the average entanglement entropy in the disordered model and the entanglement entropy for the homogeneous case.
In agreement with the magnetization reported in Fig. \ref{fig:Dimer}, $E$ exhibits a strong discontinuity at ${U_{2C}=0}$ in the homogeneous case while varying continuously in the presence of disorder.
 It is interesting to note that while the disordered system is unable to reach the entropy values obtained in the weak dimerized regime ${0<U_{2C}\ll 1}$, the disordered system has on average a nonzero entropy in the ferromagnetic phase. This confirms the results Fig. \ref{fig:Dimer} suggests: For negative $U_{2C}$ the chain is not fully magnetized and spins show strong correlations signaled by a nonzero entanglement entropy.

\begin{figure}[t] 
\centering
\includegraphics[trim=7cm 5cm 13cm 0.75cm, clip=true, width=0.475\columnwidth]{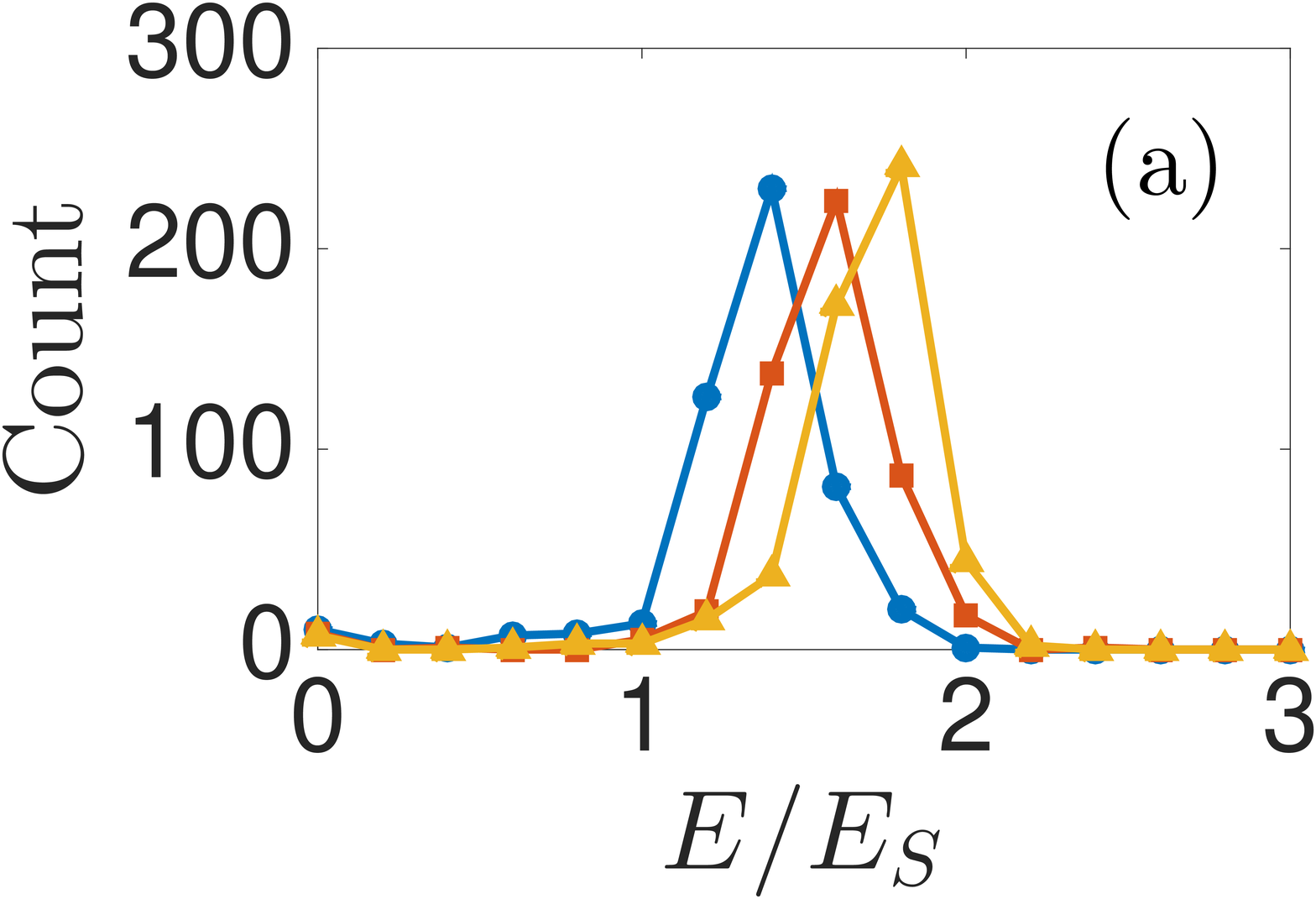}
\includegraphics[trim=11.5cm 5cm 8.5cm 0.75cm, clip=true, width=0.475\columnwidth]{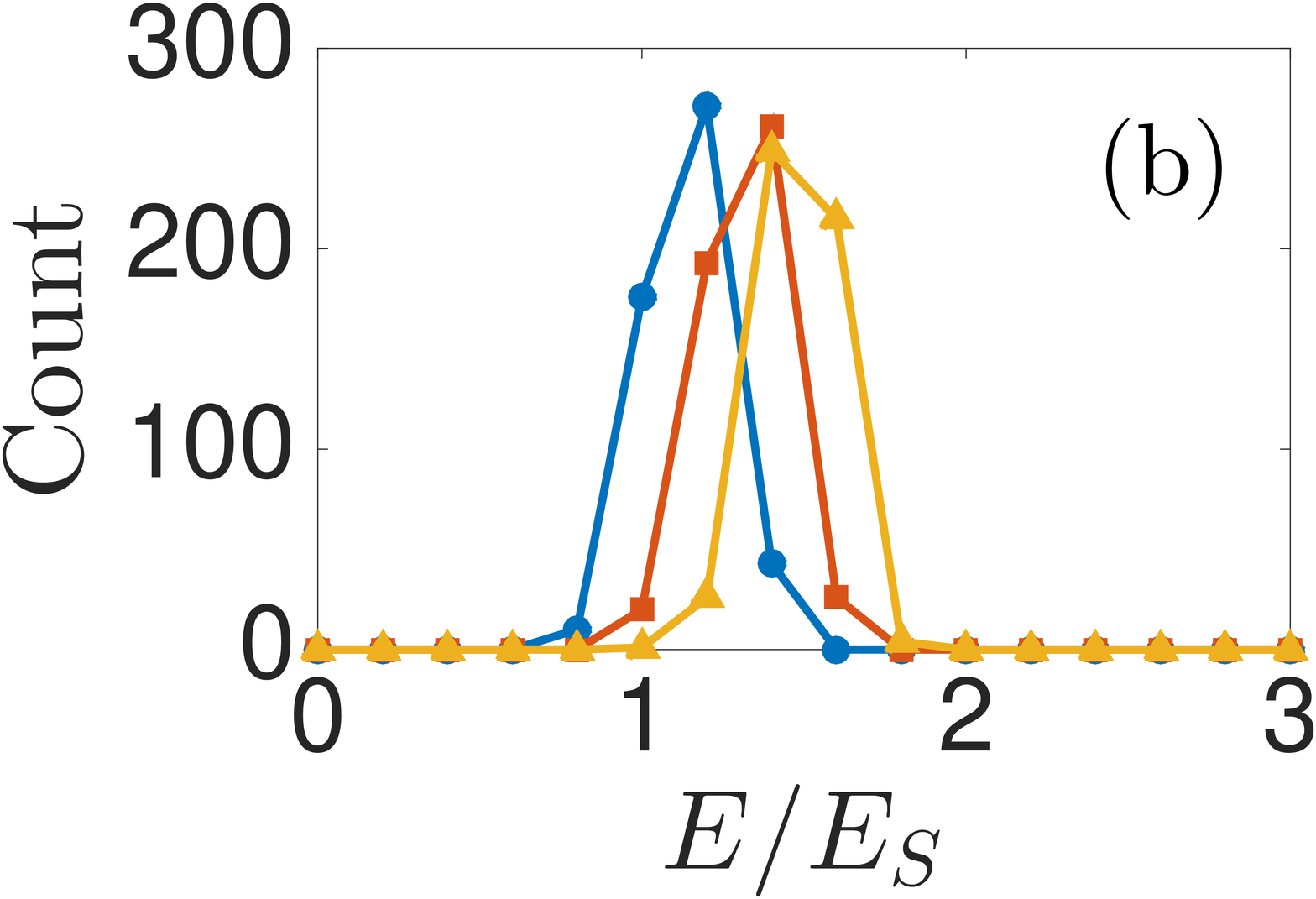}\\
\includegraphics[trim=7cm 0.75cm 13cm 0.75cm, clip=true, width=0.475\columnwidth]{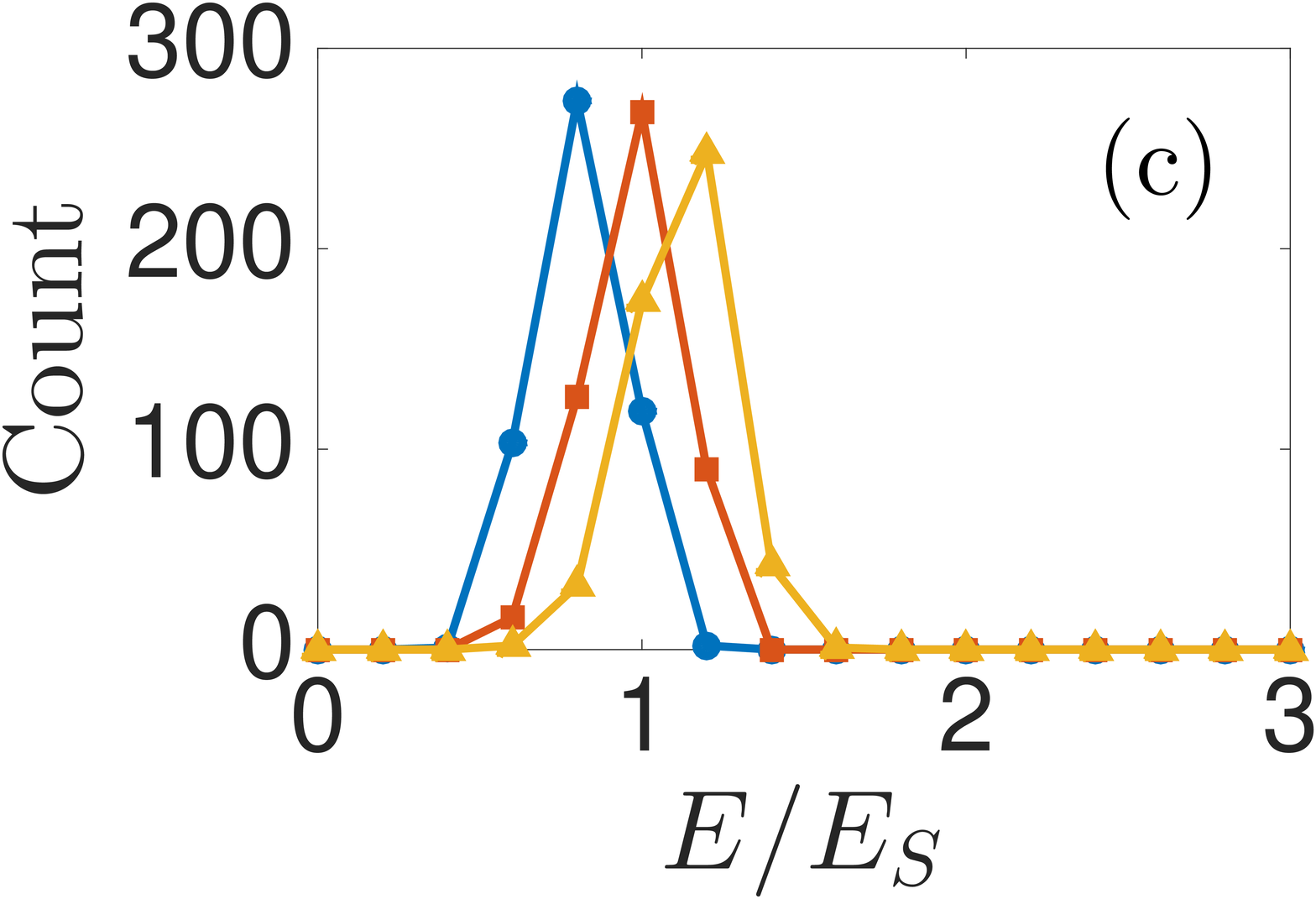}
\includegraphics[trim=11.5cm 0.75cm 8.5cm 0.75cm, clip=true, width=0.475\columnwidth]{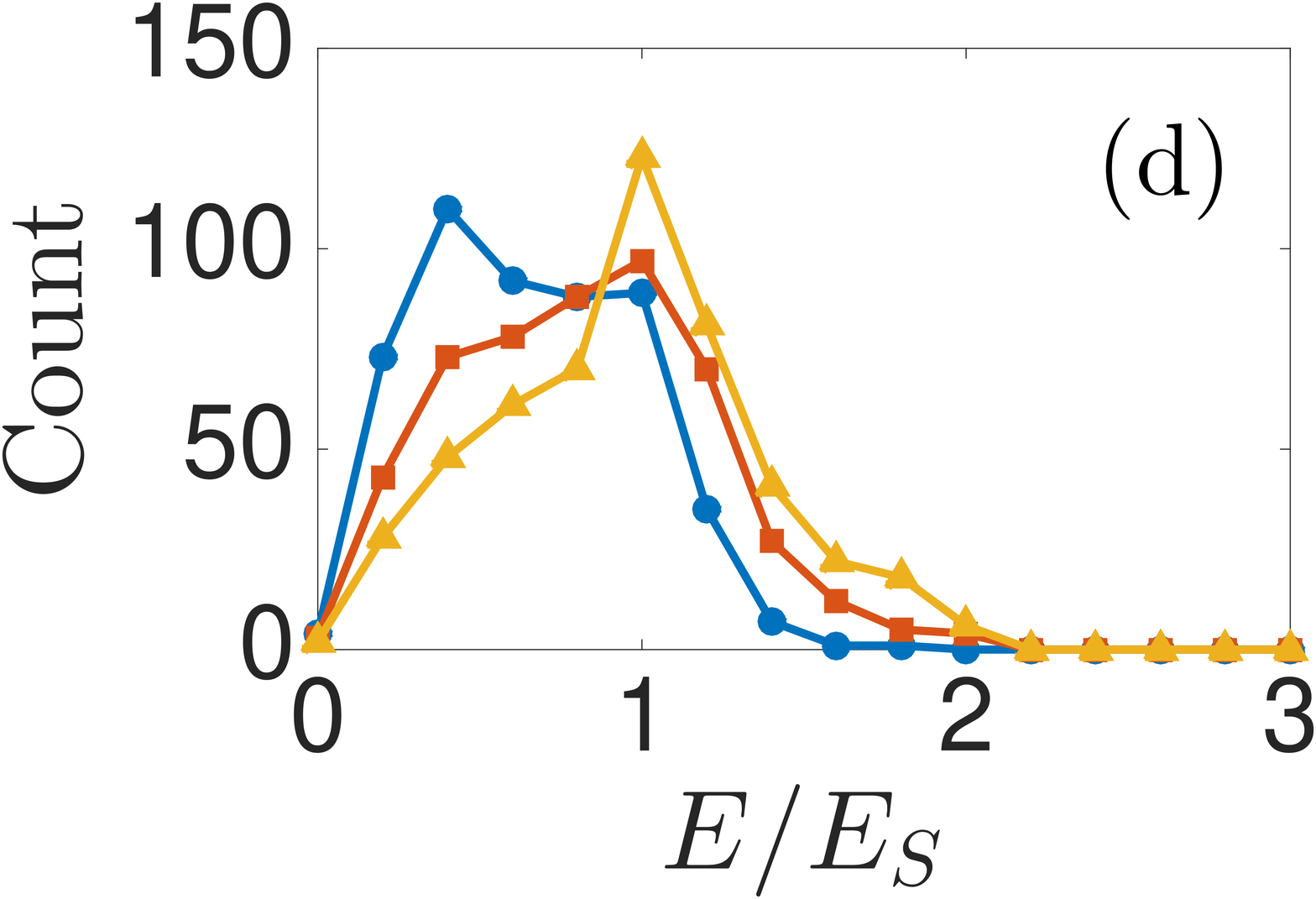}
\caption{Distribution of entanglement entropy values normalized by the entropy of a single spin-1 singlet for $U_{2C}$ values (a) 0.05, (b) 0.1, (c) 0.2, (d) and 0.39. Same color-coding as in Fig. \ref{fig:Qprime}.}
\label{fig:VNHist}
\end{figure}

In Fig. \ref{fig:VNHist}, we show histograms of the entanglement entropy for ${U_{2C}=0.05, 0.1, 0.2, 0.39}$. These values correspond in the homogeneous system to the dimer phase. Since we are interested in detecting the random singlet phase, we plot the histogram so that the horizontal axis displays the ratio $E/E_S$, where ${E_S = \log_2(3) \simeq 1.585}$ is the entanglement of a spin-1 singlet (which is maximally entangled state as for spin-1/2 particles). We find that the entanglement distribution exhibits a broad peak for ${E>E_S}$, whose midpoint decreases for increasing $U_{2C}$ and increases for increasing $L$. For ${U_{2C}=0.39}$, the distribution is much wider.
In the cases considered, we observe the major peak tending towards a value of $E_S$ as predicted for the random singlet phase for the model presented in Ref. \cite{Quito15}. However, probably because of the finite lengths considered in our simulations, we do not observe peaks occurring at other integer multiples of $E_S$.
\vspace{-12pt}
\subsection{\hspace{-6pt}Domain walls}\label{sec:coex}
\vspace{-12pt}
Thus far we have shown that, through the introduction of disorder, an intermediate phase can be observed between the ferromagnetic and dimer phases.
This phase is characterized by a nonzero EA order parameter coexisting with a nonzero magnetization and a possible exponentially small dimer order parameter.
Let us concentrate on the local magnetization properties.
From our analysis in Sec. \ref{sec:local}, we know that in the ferromagnetic phase neighboring spins are aligned, thus having ${\langle S_{zi} S_{zi+1}\rangle >0}$. Conversely, in the dimer phase, spins tend to form imperfect singlets and thus have a tendency to antialign, giving rise to a negative value of $\langle S_{zi} S_{zi+1}\rangle$. Thus in both of these phases, the sign of $\langle S_{zi} S_{zi+1}\rangle$ is constant and  by observing sign changes in its value we can detect a domain wall.

\begin{figure}[t] 
\centering
\includegraphics[trim=9cm 0.45cm 10cm 1.5cm, clip=true, width=0.95\columnwidth]{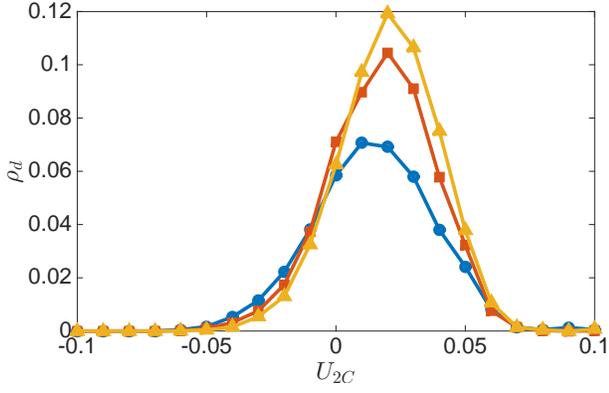}
\caption{
Average density of domain walls $\bar{\rho}_d$  against $U_{2C}$. A value of $0$ indicates that the entire spin chain is in a ground state associated with either the ferromagnetic or dimerized phase. Same color-coding as in Fig. \ref{fig:Qprime}.
}
\label{fig:Coexist}
\end{figure}

Figure \ref{fig:Coexist} shows the average density of domain walls, $\bar \rho_d$, occurring in a given chain of length $L$ for different $L$.
A thermodynamically large number of domain walls, corresponding to small droplets of magnetized spins, is found in the intermediate region with the maximum density reached around $U_{2C}\simeq 0.017$, indicating  that the dimer phase is more susceptible to disorder than the ferromagnetic phase. This could be ascribed to the much smaller gap in the dimer phase as $U_{2C}\to0^+$.
By comparing Fig. \ref{fig:Coexist} and Fig. \ref{fig:Qprime}, we see that the density of domain walls and the EA order parameter follow the same qualitative behavior. In other words, they are indeed markers of the intermediate phase. It is easy to show that within small disorder expansion, the EA order parameter is proportional to the local susceptibility with respect to  changes in the local spin-spin coupling. On the other hand, a sizable amount of domain walls measure how the ground state locally changes from antiferromagnetic to ferromagnetic short-range correlations due to different values of the spin couplings, which qualitatively explains the similar behavior of the two quantities.
\vspace{-12pt}
\section{\hspace{-6pt}Conclusions}
\label{sec:conclusions}
\vspace{-12pt}
In conclusion, we analyzed the zero-temperature phase diagram of spin-1 bosons in one-dimensional optical lattices with disordered interactions generated with uniformly distributed random scattering lengths. We found that between the ferromagnetic and dimer phases occurring in the clean case, there exists an intermediate phase showing features of a disordered ferromagnet and characterized by a finite EA order parameter.
It displays disorder-induced magnetized microscopic droplets whose concentration, as well as the EA parameter, does not show a strong dependence on the system size.
The strength of the local magnetization in each droplet is locally correlated with the disorder interactions for each site. We cannot exclude that a vanishingly small dimer order is still present in this disordered phase. If this were the case, there exists the possibility that a weak local dimer order parameter is present in the regions where the magnetization is smaller. Thus we cannot exclude the existence of a Griffiths-type phase between the intermediate phase and the random singlet phase.

We thus conjecture that this intermediate phase coincides with the large-spin phase predicted by Quito {\it et al.} \cite{Quito15} by observing that the average magnetization per particle has an intermediate value between zero and one.

The intermediate disordered phase could potentially be close to the nematic region conjectured by Chubukov \cite{Chubukov91}. It would  therefore be  interesting to analyze the stability
of the intermediate phase with respect to the appearance of the nematic  phase when a uniaxial field is added \cite{DeChiaraAug11}. This study is beyond the scope of the current paper and we leave it for future investigations.

In analogy with the classical version of the bilinear-biquadratic spin-1 model \cite{Ara2000}, which  for infinite range interactions exhibits a spin glass phase, one can ask if also in the present case the disordered ferromagnet displays some nonequilibrium properties typical of spin glasses. It remains an open problem to determine if the replica symmetry is actually broken and in order to give a precise answer a further study of the long-time dynamics would be in order. We think this is an interesting subject for a subsequent project.

\vspace{-12pt}
\acknowledgements
\vspace{-12pt}
We acknowledge helpful discussions with C. Cartwright. K.D.M. wishes to thank the Department for Employment and Learning (DEL) and the Professor Caldwell Travel Studentship for support.
S.P. is supported by a Rita Levi-Montalcini fellowship of MIUR. A.S. acknowledges financial support from the Spanish
MINECO Projects FIS2013-40627-P and FIS2016-80681-P
 and the Generalitat de Catalunya CIRIT (2014-SGR-966).

\appendix
\vspace{-12pt}
\section{DERIVATION OF THE EFFECTIVE SPIN-HAMILTONIAN} \label{sec:derivation}
\vspace{-12pt}
In this Appendix, we explain how to derive the bilinear-biquadratic Hamiltonian Eq. \eqref{Hamiltonian} from the spin-1 Bose-Hubbard Hamiltonian Eq. \eqref{Bose} in the MI phase by second-order perturbation theory with respect to the tunneling amplitude.

In the MI phase, at zero temperature and without any disorder, the zero-order ground state of \eqref{Bose} is given by the degenerate ground state of the unperturbed interaction Hamiltonian
\begin{equation}\label{eqn:H0}
H_0=\frac{U_0}{2}\sum_in_i\left(n_i-1\right)+\frac{U_2}{2}\sum_i \left(\mathbf{S}_i^2-2n_i\right)-\mu\sum_in_i
\end{equation}
given by the product state
\begin{equation}\label{eqn:psi_0}
\ket{\left\{ m \right\}}=\prod_i \ket{n_i=n;s_i=s;m_i},
\end{equation}
where $n$ and $s$ are  fixed and  the same for each site and $m_i$ is free to change. The configuration of all the $m_i$s in the lattice is denoted with $\left\{ m \right\}$ and
\begin{equation}
H_{0}\ket{\left\{ m \right\}}=E_0 \ket{\left\{ m \right\}}.
\end{equation}
 We focus on the case with  filling $n=1$ and $s=1$: A small amount of hopping does not affect the on-site density so we can construct an effective Hamiltonian  $H_{\text{eff}}$ acting only  in this subspace. To determine   $H_{\text{eff}}$, one can perform a second-order perturbation expansion in the hopping term
\begin{equation}\label{eqn:T}
T=-t\sum_{i,\sigma} \left(a_{i,\sigma}^\dagger a_{i+1,\sigma}+a_{i+1,\sigma}^\dagger a_{i,\sigma}\right),
\end{equation}
following the procedure described in Ref. \cite{cohen1992}
\begin{equation}\label{eqn:Heff}
H_{\text{eff}}=H_{0}+\sum_{\left\{ m \right\},\left\{ m \right\}'} \ket{\left\{ m \right\}'}\bra{\left\{ m \right\}}\sum_\alpha \frac{\bra{\left\{ m \right\}}T\ket{\alpha}\bra{\alpha}T\ket{\left\{ m \right\}'}}{\Delta_\alpha},
\end{equation}
where $\ket{\alpha}$ are intermediate eigenstates of $H_0$ with
\begin{equation}
H_{0}\ket{\alpha}=E_\alpha \ket{\alpha}
\end{equation}
and
\begin{equation}
\Delta_\alpha= E_0-E_\alpha.
\end{equation}

The effective Hamiltonian is expected to maintain the same SO(3) symmetry of the original model, so it must have the form of the bilinear-biquadratic Hamiltonian $H_{BB}$  \eqref{Hamiltonian} with  $J_{1i}$ and $J_{2i}$  depending on $t$, $U_0$, $U_{2i}$, and $U_{2(i+1)}$.
Following Ref. \cite{Imambekov03}, the coefficients $J_{1i}$ and $J_{2i}$  are determined by comparing $H_{BB}$ with \eqref{eqn:Heff}.

In our case, the spin interaction $U_{2i}$  is a site-dependent disordered variable, thus giving rise to  coefficients $J_{1i}$ and $J_{2i}$ dependent on the bond ${(i;i+1)}$. To determine them explicitly, we consider a two-site system (only one bond) with a vector basis $\ket{ \left\{ m \right\} }=\ket{m_1,m_2}$ and calculate the matrix elements
\begin{eqnarray}
\braket{m_1+1,m_1-1|H_{\text{eff}} |m_1,m_2} & = & D_1(m_1,m_2), \nonumber \\
\braket{m_1+2,m_1-2|H_{\text{eff}} |m_1,m_2} & = & D_2(m_1,m_2),
\end{eqnarray}
with
\begin{eqnarray}\label{eqn:ds0}
D_1(m_1,m_2) & = & \frac{W(m_1,m_2)}{2}\left[J_1+J_2  W_1(m_1,m_2)\right], \nonumber \\
D_2(m_1,m_2) & = &   \frac{J_2}{4}W(m_1,m_2)W(m_1+1,m_2-1),
\end{eqnarray}
and
\begin{eqnarray}
W(m_1,m_2)     & = &   \sqrt{(1-m_1)(2+m_1)(1+m_2)(2-m_2)}, \nonumber \\
W_1(m_1,m_2) & = &   2m_1m_2+m_2-m_1-1.
\end{eqnarray}
To our purpose, it is sufficient to consider only two elements:
\begin{eqnarray}
\braket{1,-1|H_{\text{eff}} |0,0} & = & D_1(0,0)=J_1-J_2, \nonumber \\
\braket{1,-1|H_{\text{eff}} |-1,1} & = & D_2(-1,1)=J_2.
\end{eqnarray}
For the calculation of the matrix elements, we use \eqref{eqn:Heff}, taking into account that the creation and annihilation operators act as
\begin{eqnarray*}
a^\dagger_\sigma \ket{n;s;m}&=&\bar{A} \ket{n+1;s+1;m+\sigma}+\bar{B} \ket{n+1;s-1;m+\sigma},\\
a_\sigma \ket{n;s;m}&=&A \ket{n-1;s+1;m-\sigma}+B \ket{n-1;s-1;m-\sigma}\nonumber.
\end{eqnarray*}
with coefficients $A$, $B$, $\bar{A}$, and $\bar{B}$ given in Ref. \cite{Tsuchiya04}.
There are four involved  intermediate states:
\begin{eqnarray*}
\ket{\alpha_{1,2}}&=&\ket{n_1=2;n_2=0;s_1=0,2;s_2=0;m_1=0;m_2=0},\nonumber \\
\ket{\alpha_{3,4}}&=&\ket{n_1=0,n_2=2;s_1=0:s_2=0,2;m_1=0;m_2=0},
\end{eqnarray*}
with
\begin{eqnarray}
\Delta_{\alpha_1}&=&U_0-2U_{2\,1},\nonumber \\
\Delta_{\alpha_2}&=&U_0+U_{2\,1},\nonumber \\
\Delta_{\alpha_3}&=&U_0-2U_{2\,2},,\nonumber \\
\Delta_{\alpha_4}&=&U_0+U_{2\,2},
\end{eqnarray}
\begin{widetext}
\noindent from which one obtains
\begin{eqnarray}\label{eqn:ds}
D_1(0,0) &=&\frac{2 t^2}{3}\left(\frac{1}{U_0-2
U_{2\,1}}+\frac{1}{U_0-2 U_{2\,2}}-\frac{1}{U_0+ U_{2\,1}}-\frac{1}{U_0+ U_{2\,2}}
\right), \\
D_2(-1,1) &=&-\frac{ t^2}{3} \left(\frac{2}{U_0-2
U_{2\,1}}+\frac{2}{U_0-2 U_{2\,2}}+\frac{1}{U_0+ U_{2\,1}}+\frac{1}{U_0+ U_{2\,2}} \right)\label{eqn:ds1}.
\end{eqnarray}
The same calculation can be done for every bond and, putting together \eqref{eqn:ds0}, \eqref{eqn:ds}, and \eqref{eqn:ds1}, one obtains
\begin{eqnarray*}\label{eqn:ds2}
\frac{J_{1i}}{t^2} &=&-\left(\frac{1}{U_0+ U_{2i}}+\frac{1}{U_0+ U_{2{(i+1)}}} \right), \nonumber \\
\frac{J_{2i}}{t^2} &=&-\frac{ 1}{3} \left(\frac{2}{U_0-2 U_{2i}}+\frac{2}{U_0-2 U_{2{(i+1)}}}+\frac{1}{U_0+ U_{2{i}}}+\frac{1}{U_0+ U_{2{(i+1)}}} \right).
\end{eqnarray*}

The general relations for the coefficients with fixed odd $n$ and $s=1$ are
\begin{equation}
\begin{aligned}\label{J1}
\frac{J_{1i}}{t^2}=&-\frac{(2+n)(4+n)}{15}\left(\frac{1}{U_0+U_{2i}}+\frac{1}{U_0+U_{2{(i+1)}}}\right)\\
&+\frac{4(-1+n)(4+n)}{75}\left(\frac{1}{U_0+U_{2i}+3U_{2{(i+1)}}}+\frac{1}{U_0+U_{2{(i+1)}}+3U_{2i}}\right)\\
&-\frac{(-1+n)(1+n)}{15}\left(\frac{1}{U_0-2U_{2i}+3U_{2{(i+1)}}}+\frac{1}{U_0-2U_{2{(i+1)}}+3U_{2i}}\right),
\end{aligned}
\end{equation}

\begin{equation}
\begin{aligned}\label{J2}
\frac{J_{2i}}{t^2}=&-\frac{(2+n)(4+n)}{45}\left(\frac{1}{U_0+U_{2i}}+\frac{1}{U_0+U_{2{(i+1)}}}\right)-\frac{(2+n)(1+n)}{9}\left(\frac{1}{U_0-2U_{2i}}+\frac{1}{U_0-2U_{2{(i+1)}}}\right)\\
&-\frac{(-1+n)(4+n)}{225}\left(\frac{1}{U_0+U_{2i}+3U_{2{(i+1)}}}+\frac{1}{U_0+U_{2{(i+1)}}+3U_{2i}}\right)\\
&-\frac{(-1+n)(1+n)}{45}\left(\frac{1}{U_0-2U_{2i}+3U_{2{(i+1)}}}+\frac{1}{U_0-2U_{2{(i+1)}}+3U_{2i}}\right).
\end{aligned}
\end{equation}
\end{widetext}
\vspace{-12pt}
\section{SUPPLEMENTARY FIGURES}\label{SupFigs}
\begin{figure}[t]
\centering
\includegraphics[trim=8.5cm 0.45cm 10cm 1.5cm, clip=true, width=0.95\columnwidth]{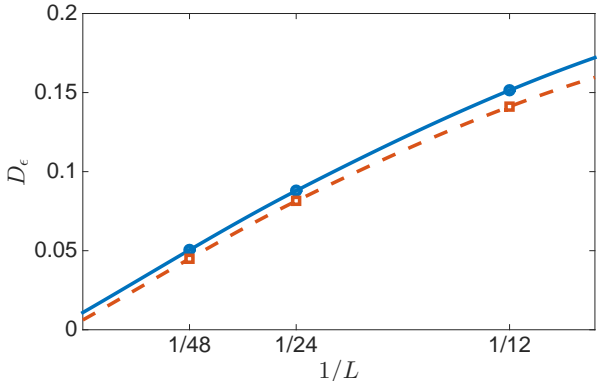}
\caption{Dimer order parameter $D_\epsilon$ extrapolation to infinite length for $U_{2C}=0.1$ for the homogeneous (circles) and disordered (squares) cases. Lines represent the extrapolating cubic polynomial fitting the data points for $L=12,24,48$.}
\label{fig:DimerExtrapU201}
\end{figure}

\vspace{-12pt}
In this section of the Appendix, we show an example of the extrapolation analysis for the dimer order parameter, mentioned in Sec. \ref{DimerSection}.
The piecewise cubic Hermite interpolating polynomial (pchip) extrapolation method in {\footnotesize MATLAB} is used to extrapolate the values of the order parameters obtained for $1/L=1/12,1/24$, and $1/48$ to the case where $1/L=0$, therefore obtaining the corresponding value for the order parameter at infinite chain length. Figure \ref{fig:DimerExtrapU201} shows how the extrapolated values of the dimer order parameter were obtained for both the homogeneous and disordered cases at $U_{2C}=0.1$.

\begin{figure}[t] 
\centering
\includegraphics[trim=8.5cm 0.45cm 10cm 1.5cm, clip=true, width=0.95\columnwidth]{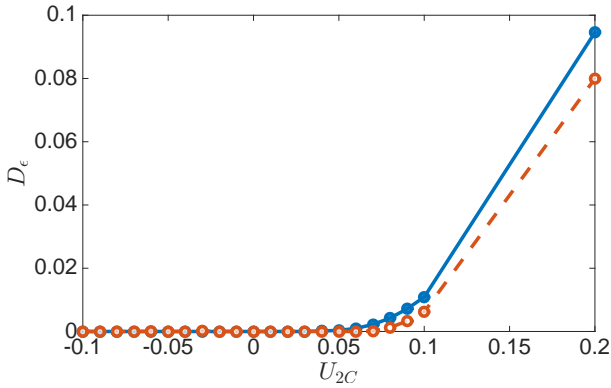}
\caption{Dimer order parameter $D_\epsilon$ against $U_{2C}$ extrapolated to infinite length. The lines (solid for homogeneous and dashed for the disordered model) connect the points and are only a guide to the eye.}
\label{fig:DimerExtrap}
\end{figure}

Figure \ref{fig:DimerExtrap} shows the results of this method when applied to the dimer order results from Fig. \ref{fig:Dimer}. These results, as previously discussed, show a significant reduction when compared to the values obtained for the dimer order parameter at finite lengths.

\bibliographystyle{apsrev4-1}
\bibliography{ReferencesBib}
\end{document}